# All-optical switching based on soliton self-trapping in dual-core high-contrast optical fibre

*M. Longobucco [a,b], J. Cimek [a,b], L. Curilla [c], D. Pysz [a], R. Buczynski [a,b], I. Bugar [a,d]*

[a] *Institute of Electronic Materials Technology, Wólczyńska 133, 01-919 Warsaw, Poland*
[b] *Department of Geophysics, Faculty of Physics, University of Warsaw, Pasteura 5, 02-093 Warsaw, Poland*
[c] *Optics11, De Boelelaan 1081, 1081 HV Amsterdam, Netherlands*
[d] *International Laser Centre, Ilkovičova 3, 841 04 Bratislava, Slovakia*

**ABSTRACT**

A systematic numerical study of ultrafast nonlinear directional coupler performance based on soliton self-trapping in a novel type of dual-core optical fibre is presented. The considered highly nonlinear fibre structure is composed of a real, intentionally developed soft glass-pair with high refractive index contrast at the level of 0.4 in the near infrared. Nonlinear propagation of picojoule level femtosecond pulses was studied numerically with the aim to identify the best switching performance in input parameter space of 1400 - 1800 nm in terms of excitation wavelengths, and of 75 - 150 fs in terms of pulse width, respectively. For every combination of excitation wavelength and pulse width, the switching energies together with the optimal fibre length were determined and their relation to the input and switching parameters is discussed. The highest switching contrast of 46 dB in the time window of the ultrashort soliton was predicted at combination of 1500 nm excitation wavelength and 75 fs pulse width considering 43 mm fibre length. These results represent significant improvement both from point of view of switching contrast and switching energies, which are only at level of 20 pJ, in comparison to the previously published case of air-glass dual-core photonic crystal fibre. Moreover, the simpler fibre design without cladding microstructure together with the all-solid approach holds promise of improved dual-core symmetry and therefore offers high probability of the successful realization of a low power, compact and simple switching device.

## 1. INTRODUCTION

Dual-core fibres represent an interesting area of applied optics, bearing potential of application as highly efficient ultrafast nonlinear directional coupler useful for signal processing and telecommunications [1]. In the dual-core fibres, the input radiation coupled in one of the two cores experiences (in the linear regime) periodic oscillations between cores, which is characterized by coupling length

$$L_c = \frac{\pi}{\beta_S - \beta_A}, \qquad (1)$$

where $\beta_S$ and $\beta_A$ are the propagation constants of symmetric and antisymmetric supermodes of the fibre, respectively, defined as the fundamental eigenmodes of a dual-core fibre and aredetermined by the fibre material and microstructure [2].

It is possible to eliminate the coupling between the cores and consequently the energy transfer between them applying intensive field, which causes significant phase changes of the in-coupled field. This is the base of the nonlinear directional coupler concept, resulting in switching from the non-excited (cross) core to the excited (bar) core when dual-core fibre with length of $L_C$ is used [3]. The concept works well in the continuous wave regime; however, it would be desirable to implement it for sub-picosecond pulses. Such approach would open the way for optical signal processing applications at Tb/s data transmission rates. However, in the ultrafast nonlinear propagation regime, the field evolution experiences complex spectral-time-spatial transformations, that hinder the switching performance potential [4,5]. In order to overcome the transformational drawbacks, it is possible to benefit from the solitonic propagation principles both in the time and spatial domain. The soliton propagation regime can preserve the pulse shape in time domain even in the dual-core fibre [1] and it can maintain asymmetric dual-core field distribution in the spatial domain in specific excitation conditions [6].

Recently a new concept was suggested, which combines the advantages of the high-order soliton self-compression with the nonlinearly induced dual-core asymmetry [7], eliminating perturbations caused by the dispersion, coupling or nonlinearity. Due to the high field intensity, which occurs after the pulse compression process, the high-order soliton tends to be self-trapped in one fibre core and to propagate further in nearly undistorted form. In such a way, by proper choice of the initial peak energies, we can create the situation where the lower energy pulse is confined in the non-excited core, while the higher energy pulse in the excited core, so nonlinear switching takes place. The concept has been introduced first theoretically by optimizing the geometry of dual-core photonic crystal fibre (PCF) made of highly nonlinear glass, which supports controllable self-trapping of high-order solitons in its both cores [7]. The numerical results obtained in this study suggested sub-nanojoule switching energies and high-contrast switching possibilities above 30 dB in the C-band, never demonstrated before in the case of fibre based nonlinear directional coupler. However, so far there was no successful experimental realization of such a scenario, because the manufactured photonic crystal fibre structure did not possessed sufficient initial dual-core symmetry, therefore the coupling efficiency between the cores was not high enough [8]. It is possible to overcome this limitation by taking advantage of a new all-solid PCF technology, which avoids air holes in the fibre structure and therefore supports higher level of the structural symmetry.

The systematic optimization process considering all-solid dual-core (DC) PCF made of two thermally matched high index contrast glasses is the main subject of this paper. Theoretical considerations regarding the fibre characteristics, that enable the advantageous switching, will be introduced in the second section, together with the numerical methods used for the study of the linear and nonlinear fibre properties. Description of the optimized fibre structure with its basic linear spectral characteristics will be presented in the third section. The fourth section will be dedicated to nonlinear solitonic switching study through the optimized dual-core fibre by numerical solution of the coupled generalized nonlinear Schrödinger equations. Excitation wavelength and pulse width effect on the switching performance will be discussed in details and the best results will be compared to the previously published numerical study performed on air-glass DC PCF [7]. Realization perspectives of the presented new fibre structure together with its application potential will be discussed at the end of the paper.

## 2. PREDISPOSITIONS

### 2.1 Theoretical background

Optical solitons can oscillate between the two fibre cores without distortion, like CW radiation, and can support effective nonlinear switching induced by the change of the input energy (Figure 1a.) [1]. However, this theoretical concept has never been demonstrated experimentally, because the soliton-like pulses in fibres are exposed to various distortions, mainly due to third order dispersion (TOD) and stimulated Raman scattering (SRS). Moreover the coupling coefficient dispersion causes additional pulse brake-up in the case of dual-core fibres [9,10]. On the other hand, when we move from the weak nonlinearity to the strong nonlinearity regime, it is possible to avoid these drawbacks. It is generally known, that when the nonlinear length is equal to the dispersion length, first order soliton effect occurs, preventing the pulse brake-up in the time domain. Analogously, one can expect, that when the nonlinear length is of the same order as the coupling length the pulse brake-up can be suppressed also in the spatial domain. Our previous numerical study considering an air-glass DC PCF confirmed this idea, uncovering the effect of self-trapped solitonic propagation in the case significant self-compression of high-order solitons [7]. Due to high peak power after the pulse compression process and subsequently induced strong nonlinear asymmetry between the two cores, the soliton-like pulse is self-trapped in one fibre core and propagates further nearly undistorted. At proper choice of the pulse input energy, which was in the case of the cited study in the sub-nanojoule range, it is possible to exchange the core of the field confinement with high switching contrast (Figure 1b.).

The condition for the soliton self-trapping is obtained when we balance soliton compression distance and the coupling length. To evaluate the former it's necessary to introduce the soliton number (order), defined as [see ref.11]:

$$N^2 = \frac{\gamma P_0 T_0^2}{|\beta_2|}, \quad (2)$$

where $P_0$ is the pulse peak power, $\gamma$ is the nonlinear coefficient, $T_0$ is the pulse duration and $\beta_2$ is the group velocity dispersion. When $N > 2$, a higher-order soliton is generated, which periodically recovers after a soliton period $z_0$, defined as a function of pulse duration and group velocity dispersion expressed by relation [11]

$$z_0 = \frac{\pi T_0^2}{2|\beta_2|}. \quad (3)$$

Due to high-order perturbation effects the high-order soliton undergoes compression and subsequent soliton fission process. It results in a broadband output spectra, at the distance $z_{comp}$, which, for $N > 10$, can be approximated using equation [2]

$$z_{comp} = \left(\frac{0.32}{N} + \frac{1.1}{N^2}\right) z_0 \qquad (4)$$

According to the formulae (2) - (4), the initial compression distance $z_{comp}$ gradually decreases from the soliton period value with increasing peak power (soliton order), because $z_0$ does not depend on it.

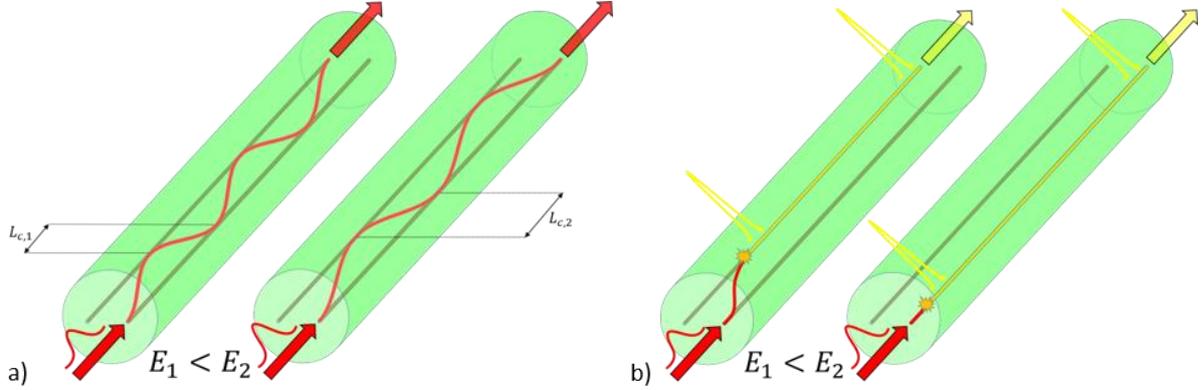

*Figure 1. a) Solitonic propagation through a dual-core fibre in weak nonlinear regime: the radiation in-coupled into one of the cores oscillates between the two channels with coupling length $L_c$, whose value is power-dependent. b) Theoretical concept of self-trapped solitonic switching: after the pulse compression phase the high peak powers trap the shortened pulse in the cross or in the bar core, which choice is controllable by the input pulse energy. The pulse propagates further in the same core in quasi-stable conditions.*

When $z_{comp}$ is tuned by the pulse energy slightly above $L_c$, the high-order soliton is compressed just after its transfer to the cross core. In contrast, the sufficiently higher pulse energy (due to the shorter $z_{comp}$) causes compression maxima already in the bar core. In the both cases, the high peak powers of the compressed pulses induce strong phase mismatch between the cores and in this way, the cores are effectively decoupled during the further propagation. For these reasons, the key condition of the self-trapped solitonic switching concept is to achieve the comparable fibre coupling length and soliton compression length in certain spectral band. We will focus mainly on the C-band of the optical communication systems. In order to offer applicable switching performance, the few mm-cm coupling length region is targeted, representing good compromise between the compactness and easy handling aspects.

Further conditions of the fibre structure optimization are: maintain the pulse energies in the picojoule region and eliminate the energy loss of the main soliton during the soliton fission process. The first condition is ensured by using highly nonlinear glass PBG-08 for the fibre glass, which exceeds 20 times the nonlinear refraction index of fused silica. By this choice, simple ultrafast fibre oscillators with pulse durations at the level of 100 fs are applicable for the future experimental realization. The second condition is ensured by keeping the soliton number below value of 5, what is manageable by moderate level of anomalous dispersion considering 100 pJ pulse energies with duration of 100 fs, according to equation (2). Finally, in order to eliminate the dispersive wave generation effect, an attention was paid to reduce the TOD during the geometry optimization process.

**2.2 Numerical methods**

The structure of the dual-core fibre was initially optimized by the proper choice of material used for manufacturing fibers. The choice of the two glasses used for the cores or for the cladding was a result of an extensive research, including synthesis of a complementary glass to the highly nonlinear lead–bismuth-gallium-silicate (PBG-08) glass, used already earlier also in the case of the air-glass DC PCF structure [7,8]. The precise composition of the complementary borosilicate type glass was tuned in order to match the thermo-mechanical properties of the high refraction index, guiding PBG-08 glass. The optimization process resulted in glass composition labelled as UV-710 allowing air-like behaviour of the areas filled with it at refraction index contrast around 0.4 in a wide spectral region (1400 - 2000 nm) [12]. The main achievements of the material optimization can be summarized as follows:

- Low thermal stress due to small difference in thermal expansion coefficient ($3.9 \cdot 10^{-6}$ K$^{-1}$) between the two glasses resulting in sufficient thermal stability;
- No air pressure fluctuations: due to all-solid approach, the fibre has more homogeneous structure with respect to the air-glass case;

- High transmission window between 600 and 2650 nm;
- High refractive index constant difference (0.422 at 1500 nm) enabling to ensure significant anomalous dispersion.

At the beginning, a periodic hexagonal lattice of hexagonal low index glass rods was considered (Figure 2a.), an approach which was already successfully implemented in the case of single core all-solid PCFs [13]. Dual-core fibre, which avoids completely the PBG-08 glass bridges between UV710-glass hexagonal rods (Figure 2b.) was considered as an alternative. The latter offers the maximum reachable low index fill factor of the cladding and significant simplification of the manufacturing process.

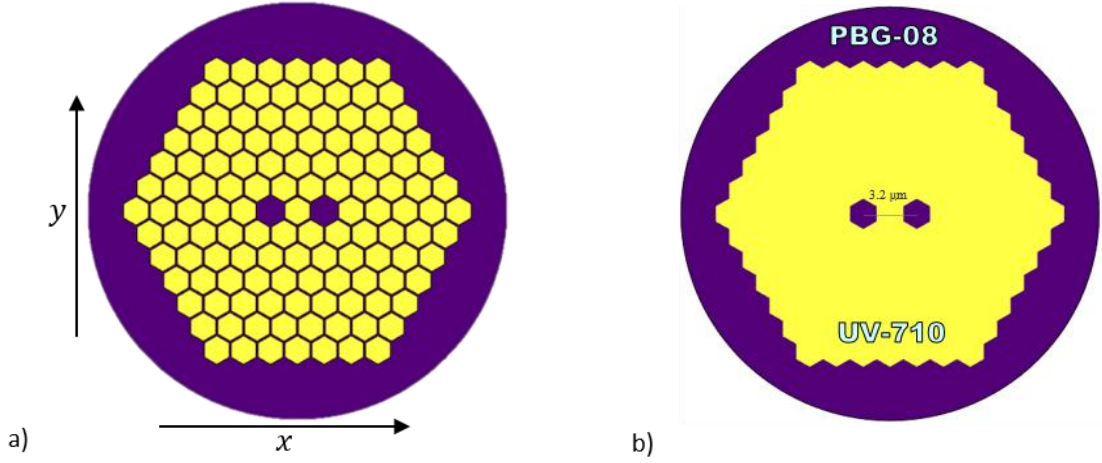

*Figure 2. Hexagonal array of hexagonal rods type of a dual-core all-solid PCF fibre with lattice with (a) high and (b) maximal low index fill factor of the cladding. In the latter case, the two cores and the low index microrods have the same diameter.*

In the next step, the new structure was studied from the point of view of optical field propagation characteristics in linear regime. The commercial Mode Solution software from Lumerical was used to calculate the spectral dependences of the field mode profile, the corresponding effective index and the waveguide losses for each fundamental mode. The all relevant quantities like effective mode area, dispersion, coupling coefficient, losses were acquired in spectral window between 0.3 and 4.1 μm, which covers sufficiently our region of interest. Both mode identification and dispersion characteristics calculation were initially performed for a single core structure, that was obtained by filling one of the cores with low index glass. In the second step, the more complex dual-core structure was used to determine the coupling coefficient between the two single core modes based on the overlap integrals [2]. It's important to mention, in the case of a dual-core fibre, a set of four fundamental supermodes (symmetric and anti-symmetric with both X and Y polarization directions) is obtained, while a set of two orthogonally polarized fundamental modes are considered in the case of a single core fibre.

The single core fibre characteristics together with the identified coupling coefficients are then used in the nonlinear simulation, where the coupled generalized nonlinear Schrödinger equations (CGNSE) are solved numerically. The form of these equations was adapted to the case of dual-core PCF by incorporating effects like coupling coefficient dispersion, self-steepening nonlinearity and its spectral dependence, stimulated Raman contribution, cross-phase modulation effect and waveguide losses. The resulting mathematical model is a system of two equations expressed in the retarded time frame $T$ in the following form [5]:

$$\frac{\partial A_r}{\partial z} = (-1)^{r+1}i\delta_0 A_r + (-1)^r \delta_1 \frac{\partial A_r}{\partial T} - \frac{1}{2}\sum_{o\geq 0}\frac{i^o}{o!}\alpha_o^{(r)}\frac{\partial^o A_r}{\partial T^o} + \sum_{p\geq 2}\frac{i^{p+1}}{p!}\beta_p^{(r)}\frac{\partial^p A_r}{\partial T^p} + \sum_{q\geq 0}\frac{i^{q+1}}{q!}k_q^{(r)}\frac{\partial^q A_{3-r}}{\partial T^q} + \\ + i\gamma^{(r)}\left[\left(1 + i\tau_{shk}^{(r)}\frac{\partial}{\partial T}\right)\int_{-\infty}^{\infty}R(T')\times |A_r(z, T-T')|^2 dT' + \sigma^{(r)}|A_{3-r}|^2\right]A_r. \quad (5)$$

Here $r = 1, 2$ denotes the number of the cores, $A_r$ is the corresponding electric field amplitude and quantities $\delta_0 = (\beta_0^{(r)} - \beta_0^{(3-r)})/2$ and $\delta_1 = (\beta_1^{(r)} - \beta_2^{(3-r)})/2$ represent the difference between the phase and group velocities. Further $\alpha_k^{(r)}, \beta_k^{(r)}, \kappa_k^{(r)}$ are the $k$-th order of Taylor expansion coefficients of losses, propagation constant and coupling coefficient around the central frequency, respectively. Finally, $\gamma^{(r)}$ is the nonlinear parameter, $\tau_{shk}^{(r)}$ is the characteristic time of shock wave formation, $R$ is the Raman response function and $\sigma^{(r)}$ is the overlap integral between the single core modes defining for the cross-phase modulation effect in the $r$-th core. Both experimentally determined instantaneous Kerr and delayed Raman response of the guiding PBG-08 glass are included into the material nonlinear response function [14].

The equations were solved numerically by the split-step Fourier method with number of steps 40000, which value was already optimized during the previous air-glass PCF study. After every 200 calculations step, the field arrays were saved and then used to generate the output propagation maps; this means that the whole propagation distance is divided into 200 points. This approach represents a good compromise between the calculation time and the resolution of the propagation distance. In the case of 10 cm fibre length, the optimal switching length can be determined with a precision of 0.5 mm, which is the experimentally addressable precision. Using the split-step method, the losses, propagation constant and coupling coefficient are applied always at the frequency step, thus it considers their whole spectral behaviour. During the numerical simulations, the input pulse shape is approximated by the sech$^2$ function, which is good approximation for ultrafast oscillators with pulse widths at the level of 100 fs. The power envelope of the pulse is expressed as:

$$P(t) = \frac{0.88\, E \cdot k_{IN}}{T_{\text{FWHM}}} \text{sech}^2\left(\frac{t}{T_{\text{FWHM}}} 1.763\right), \quad (6)$$

where $k_{IN}$ is the in-coupling efficiency. During our systematic simulations, the pulse width $T_{\text{FWHM}}$ was set between 75 fs and 150 fs and the excitation wavelength between 1400 nm and 1800 nm, entirely covering the S, C, L, U - bands of infrared optical communication systems also with some margins.

## 3. CALCULATION OF BASIC FIBRE CHARACTERISTICS

After choosing the pair of the soft glasses, the structure of a dual-core fibre was optimized from the point of view of geometry, considering the following requirements:

- As high as possible anomalous dispersion in area of 1400 – 1800 nm, with particular attention to the C-band;
- Sufficient nonlinearity for picojoule pulse energies at pulse duration level of 100 fs;
- Coupling length of few millimeters/centimeters;
- Minimization of TOD effect, obtaining as flat as possible dispersion curve in the spectral range of interest.

To match this set of requirements, a structure of hexagonal shape of low index areas were optimized (Figure 2a.) and 1.6 μm pitch at fill factor above 90 % were identified as best from point of view of dispersion characteristics [15]. Further improvement of both linear and nonlinear properties was obtained by completely avoiding the PBG-08 glass bridges between UV-710 glass hexagonal rods keeping the optimized 1.6 μm pitch (Figure 2b.). By this new approach, the two cores and also the low index microrods have the same outer diameter of 1.85 μm and due to fill factor of 100 % even higher values of anomalous dispersion in the 1400 – 1800 nm spectral range were achieved. In this case the fabrication process, which includes the preform stacking and drawing phases, is considerably simplified and therefore higher level of structural symmetry is expected.

Due to the fact, that the nonlinear simulation code is based on coupled two generalized nonlinear Schrödinger equations, at first the linear simulation analysis was focused on the single core modes. As a consequence of the structural symmetry, the all single-core characteristics, including dispersion, are the same for both cores. In Figure 3. the dispersion curves of the fundamental modes of the both single (a) and dual (b) core high index contrast all-solid fibre are presented. As it was mentioned above, simple low index cladding was considered with outer diameter of hexagonal core of 1.85 μm and with lattice pitch of 1.6 μm. The dispersion curves have two zero dispersion points at 1280 nm and 2360 nm and the anomalous dispersion value is above 50 ps/nm/km, nearly in the whole region of interest. The dispersion maxima are at wavelength 1730 nm, resulting in only moderate TOD effect in the C-band. The polarization effect is negligible: the Y-polarized field dispersion differs from the X-polarized one just at longer wavelengths with slightly lower values beyond 2000 nm. Most importantly, the dispersion profile represents clear improvement in comparison to the high-fill factor all-solid photonic structures both in terms of anomalous dispersion value and TOD elimination [16]. The effective single-core fundamental mode area is about 1.7 μm$^2$ in the C-band, what is good compromise between the fibre nonlinearity and excitability. The latter starts to be difficult around values 1 μm$^2$ and causes high in-coupling losses.

In Figure 3b., the dispersion profiles of the 4 fundamental supermodes, which represent the combinations of the two polarizations directions (X and Y) and the two dual-core symmetry states (symmetric and antisymmetric), are introduced. The single core dispersion curves are identical to the supermodes curves at short wavelengths and they are situated between the symmetric and the antisymmetric curves (with the corresponding polarization direction) at the long wavelengths (Figure 3a.). In Figure 4., the coupling length characteristics of both orthogonal polarisations calculated according to the formula (1) is presented. Again, the polarization dependence is relatively weak, therefore in the following discussion we will focus only on the X polarisation results, because the Y polarisation ones are very similar. The simple dual-core structure express relatively long coupling length with value 2.4 cm at 1550 nm, therefore it doesn't support sub-cm length switching devices. On the other hand, the increase of $L_c$ in comparison to the air-glass DC PCF fibre (about 6 times [7]) relaxes the pulse energy requirement, because the soliton compression length increases with decreasing energy (peak power), when other quantities in equations (2) - (4) are comparable.

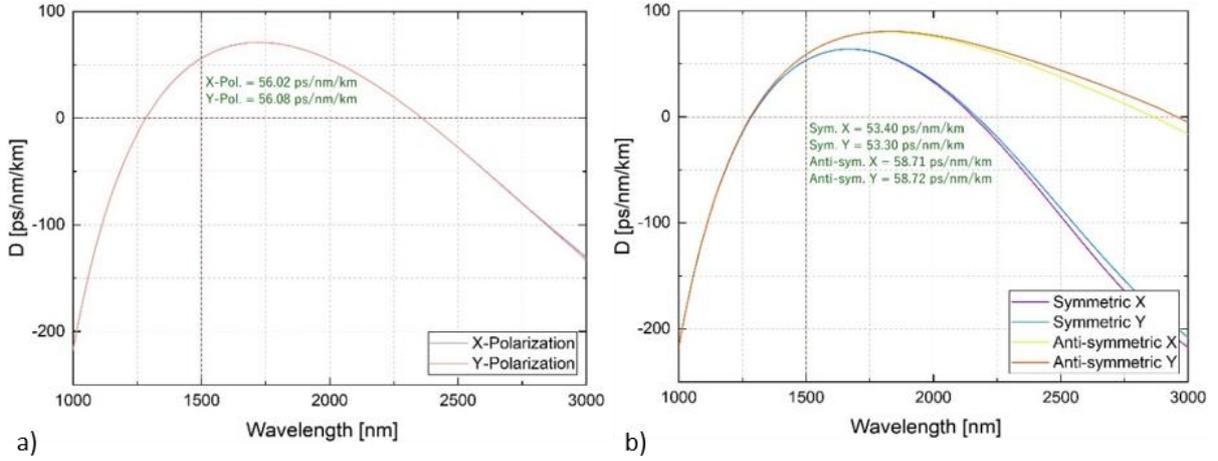

*Figure 3. Numerically calculated dispersion curves of the high-contrast all-solid fibre with 1.6 μm lattice pitch and with hexagonal core shape of 1.85 μm outer diameter: a) fundamental X and Y-polarized modes of the single core structure; b) four fundamental supermodes of the dual-core structure.*

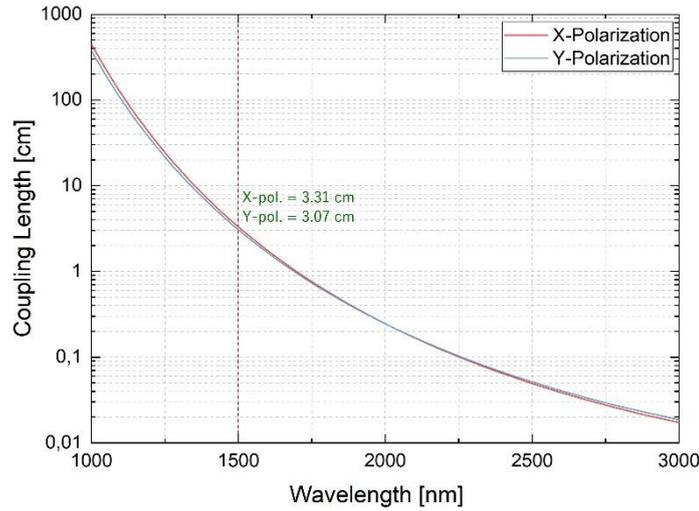

*Figure 4. Numerically calculated coupling length spectral characteristics for fundamental X and Y polarized modes of the optimized dual-core fibre. The values are calculated according to equation (1).*

Other parameters of the X polarized single core mode are the dispersion parameter $D = 62$ ps/nm/km and the effective mode area $A_{eff} = 1.72$ μm² at 1550 nm. For the further considerations, we will take these values and set the pulse width $T_{FWHM} = 100$ fs in order to create the same excitation conditions as we used in the previous work [7]. In the next step, we estimate the range of pulse energy, which supports the self-trapped solitonic switching, when condition $L_c \approx z_{comp}$ is satisfied. For this cause the dependence of input pulse peak power $P_0$ on soliton compression length can be obtained by inserting of the equations (2) and (3) for $N$ and $z_0$, into the soliton compression length equation (4). After easy simplifications, the following formula reads:

$$z_{comp} = 0.32 \cdot \frac{\pi T_0}{2\sqrt{\gamma P_0 |\beta_2|}} + 1.1 \cdot \frac{\pi}{2\gamma P_0}. \tag{7}$$

Furthermore, we set $|\beta_2| = \lambda^2 D/2\pi c$, where $c$ is light velocity, and $\gamma = 2\pi n_2/\lambda A_{eff}$, where $n_2$ is nonlinear refractive index of the guiding PBG-08 glass equal to $4.3 \cdot 10^{-19} m^2/W$ [14]. In order to estimate the pulse energy before the fibre in-coupling process, we will use 32 % in-coupling efficiency, which corresponds to the measured value in the case of the air-glass PCF having core from the same glass [8]. Therefore, for the sech[2] pulse shape we get, for in-coupled pulse peak power for pulse energy $E$

$$P_0 = \frac{0.88 \cdot E \cdot 0.32}{T_{\text{FWHM}}}. \tag{8}$$

$E$ was calculated using the above-mentioned relations and condition $z_{\text{comp}} = 2.4$ cm (coupling length) and we get the value of about 41 pJ. Hence one can expect, that input pulses with such energy will experience stable soliton self-trapping. It holds a promise of significant improvement in comparison to the results of simulation obtained using the air-glass PCF with 300-400 pJ energies before the in-coupling process in the case of high contrast switching performance. This estimated value will be verified in the next section, where the detailed analysis of the nonlinear propagation in the optimized dual-core all-solid structure will be presented.

## 4. NONLINEAR SIMULATION RESULTS

Many physical effects can be switched on and off in the numerical solution of the CGNSE (equation (5)) by setting the value of the corresponding coefficients equal to zero or to realistic nonzero values. In particular, we can switch on or off the individual terms like: optical losses, stimulated Raman scattering, cross-phase modulation (XPM) and spectral dependence of the optical shock wave formation. Two series of simulations for the cases: 1. all terms switched off and 2. all terms switched on, were performed and the results were compared. According to our results, incorporating all mentioned effects has a weak influence on the output fields both in time and spectral domain. We conclude that any of the mentioned extensions of the basic CGNSE equation does not change the propagation character including the self-trapped soliton-like evolution phenomena. These effects are important only for estimation of the switching energies with precision at the level of 1 pJ. On the other hand, the full simulation doesn't take much more calculation time, therefore we kept the terms switched on during the whole extensive simulation work. We only neglect the loss term, because it does not make any change in the propagation maps in the targeted range of fibre lengths of several cm.

Series of nonlinear simulations based on the presented model were performed for X polarized light at excitation wavelengths in range of 1400 - 1800 nm with 100 nm step and at pulse widths in the range of 75 - 150 fs with 25 fs step. At every combination of wavelength and pulse width, spectral and time domain series of propagation maps were generated by the numerical code in order to explore the influence of the increasing input pulse energy in range of 20 - 400 pJ with 5 pJ step. The aim of this study was to identify the lowest possible energies, for which the soliton self-trapping mechanism occurs in the non-excited and then in the excited core. Next we performed series of simulation focused on the region where the two above mentioned pulse energies were identified, with the step of 1 pJ, in order to achieve as high as possible switching contrast at the shortest possible distance from the fibre input. Figure 5. shows an example of the spectral propagation maps (in dB scale) for two input energies and in the both fibre cores at $\lambda = 1550$ nm and $T_{\text{FWHM}} = 100$ fs, representing interesting switching performance. In this case, the same excitation parameters were used, as those reported in [7] where an air-glass PCF was considered. The lowest excitation energy leading to soliton self-trapping in the non-excited core was identified as 100 pJ, while the same behaviour occurred in the excited core for energies close to 112 pJ. Simulated propagation length was fixed at 10 cm to monitor the field evolution also beyond the soliton compression point, which was in our case was about 2 cm. In Figure 5 results are presented in the form of 3D colour maps. The field intensity is presented in a logarithmic colour scale. The logarithmic colour scale is limited to -40 dB so the ultralow intensity features have uniform dark blue colour.

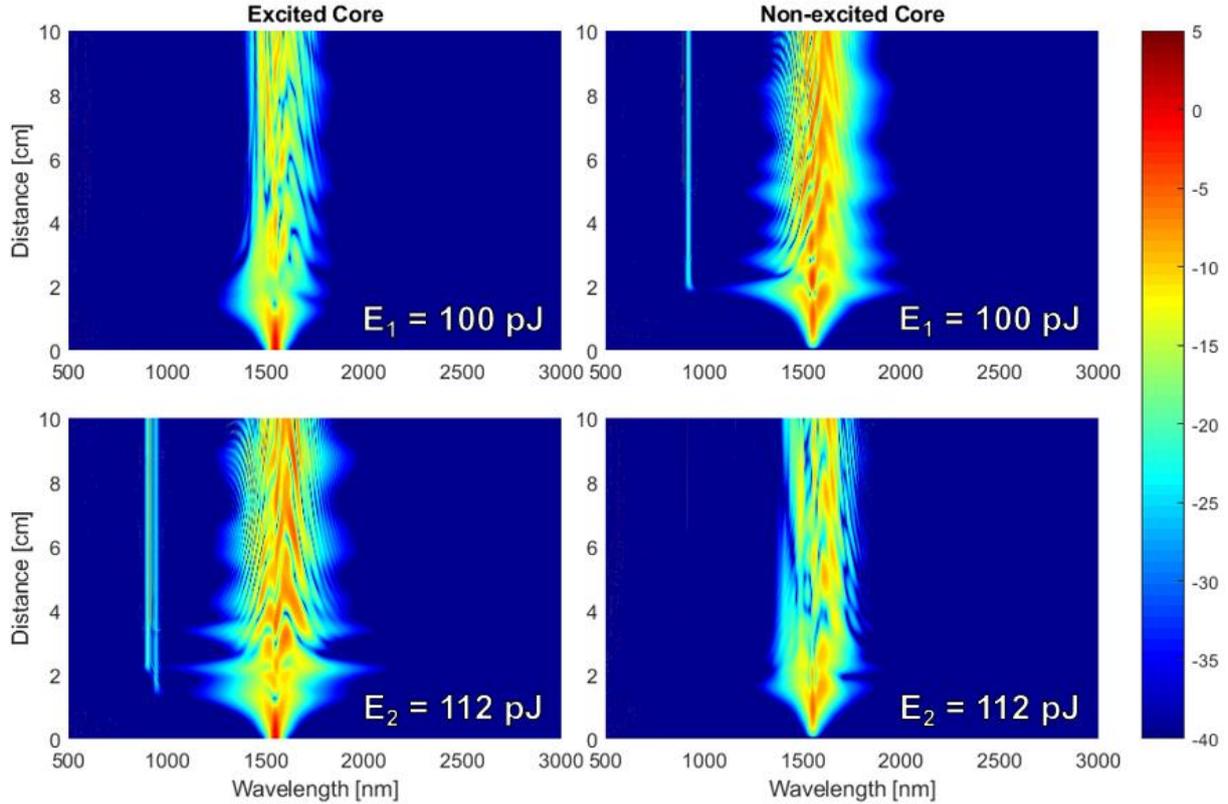

*Figure 5. Numerically generated spectral intensity maps of solitonic dual-core propagation. Excited (bar) core is presented on the left, the non-excited (cross) on the right side, respectively, under excitation by X polarized, 100 fs, 1550 nm pulses at two selected energy levels: 100 pJ (top), 112 pJ (bottom).*

The unbalanced field distribution in the bar and cross core are clearly seen in the maps at the 100 pJ energy (top row). After propagation distance of 2.4 cm, equal to the coupling length, the initial excitation in the bar core (left) is transferred to the cross one (right). It is the standard linear propagation scenario. However, the spectrum is significantly broadened in the cross core and starting from this point the dominancy of the cross core is maintained, while the bar core spectrum remains narrow. Only small amount of the pulse energy is transferred back to the bar core beyond 3 cm. It is an evident nonlinear propagation effect. The situation is different at the 112 pJ energy level: the broadest spectral feature occurs in the bar core around 2.3 cm and its remains to be dominant over whole simulated length. A periodic oscillation of the spectral width is a typical effect of high-order soliton propagation [11]. It is also confirmed by the structured spectra around the central wavelength and by the sharp spectral features in the normal dispersion region around 900 nm (so called dispersive waves). [17,18]. However, from point of view of all-optical switching, the most interesting character of the simulated maps is that the soliton-like structure continues to propagate predominantly in the core where the broadest spectrum is formed during the soliton compression part of the propagation. This is a typical signature of the soliton self-trapping mechanism with bistable character regarding the core selection, determined by the input pulse energy.

The corresponding retarded time frame intensity propagation maps (Figure 6.) show the soliton self-trapping mechanism in a more straightforward way. In the case of low-energy pulses (top row), the high-order soliton is transferred from the bar to the cross core (non-excited initially) and simultaneously its time envelope is compressed. The generated high peak powers prevent such a shortened pulse from tunnelling back to the bar core, and only partial transfer is predicted during the subsequent soliton stretching process. In the case of the high-energy pulses (bottom row), the high-order soliton is compressed at a shorter distance and therefore it is trapped already in the bar core (excited initially), keeping it there up to the end of the simulated length. Only during the first coupling length there is essential amount of energy transferred to the cross core, which after its return is trapped by the meantime shortened soliton in the bar core. The important results were to notice that the soliton compression length must be shorter than $L_C$ in order to eliminate the coupling process and to maintain the self-trapping character for longer propagation distances. It is also evident, that the higher pulse energy (112 pJ) corresponds to 1.5 cm compression length and traps more effectively the high-order soliton in the bar core, than the low energy one (100 pJ) with 2 cm compression length in the cross core. At the end of the previous section, we estimated the minimum energy required for the self-trapping process at the level of 41 pJ, when exact balance between the compression length and coupling length (2.4 cm at 1550 nm) was supposed. However, the more detailed simulation analysis revealed, that shorter compression lengths are required in order to establish the switching performance based on self-trapped solitonic propagation. According to equation (7), the shorter compression length requires higher pulse energy (peak power), than the previously estimated value. Furthermore, the energy transfer splits the pulse energy between the two cores and therefore slows down the self-compression process. The decrease of the peak power localized in one core results in additional increase of the minimum energy necessary to set up compression length at the level of 2 cm. Thus, taking into account only a single core fibre without energy transfer and longer 2.4 cm compression length, 41 pJ is a reasonable estimate. Further, it is important to emphasise, the value of 100 pJ, obtained by the rigorous numerical simulations, is just the energy before the in-coupling optics. Therefore, assuming an in-coupling efficiency of 32 %, in-fibre pulse energies are required only at the level of 32 pJ. It represents an improvement by factor of 3 to 4 in comparison to the previous theoretical air-glass PCF study [7].

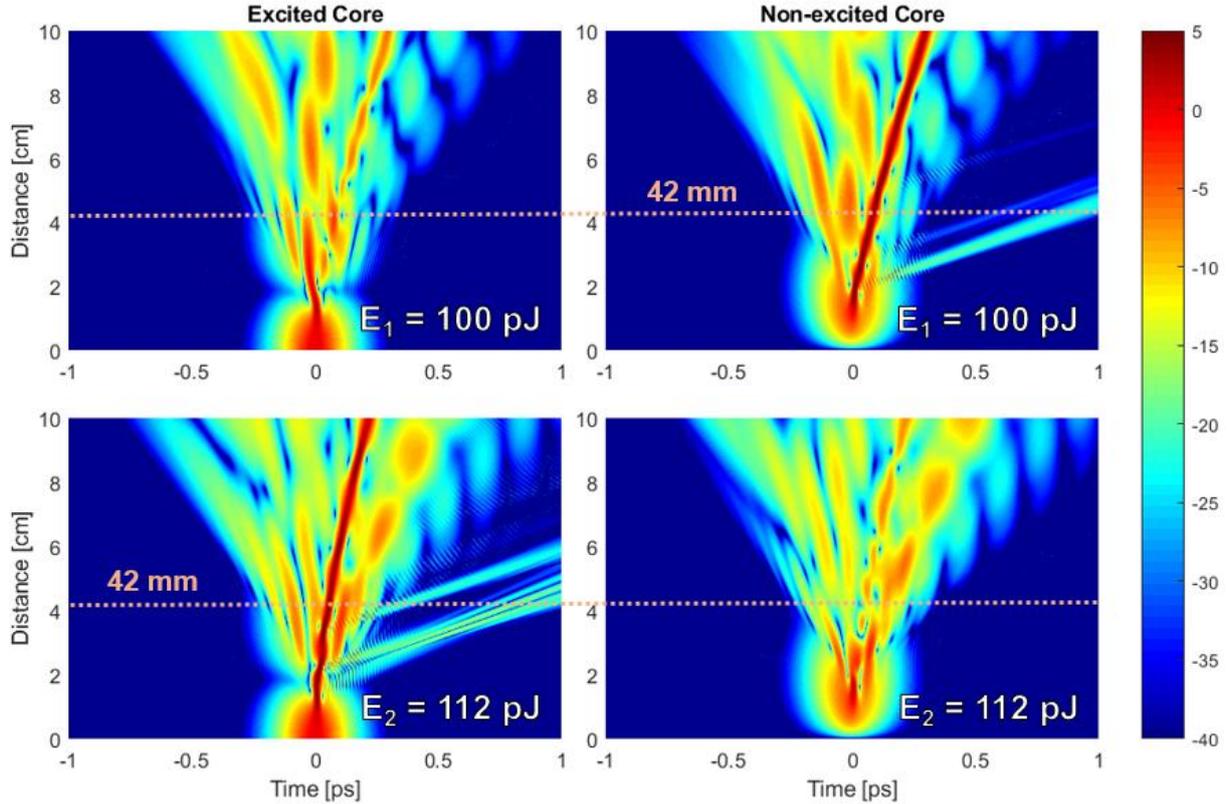

*Figure 6. Numerically generated retarded time frame intensity maps of solitonic dual-core propagation. Excited (bar) core is presented on the left, the non-excited (cross) on the right side, respectively, under excitation by X polarized, 100 fs, 1550 nm pulses for two selected energy levels: 100 pJ (top), 112 pJ (bottom).*

Next the detailed switching analysis was performed tracing carefully the temporal evolution simultaneously in the both cores along the whole simulated propagation length. As we already mentioned, the 10 cm distance is discretized by $i = 200$ steps and the integral field energies, $E_{bar}$ and $E_{cross}$ were calculated at every step in both cores. The dual-core energy extinction ratio was determined in dB according to formula

$$ER(E_{\text{input}}, z_i) = 10 log \left( \frac{E_{bar}(E_{\text{input}}, z_i)}{E_{cross}(E_{\text{input}}, z_i)} \right), \tag{9}$$

where $z_i$ is the series of discretized positions along the simulated fibre length. The most important parameter is the difference between the $ER$ in the case of the two previously identified input pulse energies, $E_1$ and $E_2$, so called switching contrast. It characterises the switching performance. The maximal switching contrast in terms of $z_i$ dependence is determined and the acquired value $z_{max}$ represents the optimal fibre length. Our analysis gives extinction ratio $ER(E_1) = -4.7$ dB with negative sign and $ER(E_2) = 7.7$ dB with positive sign, meaning dominancy of the cross and bar core, respectively. The difference between these two values is the switching contrast resulting in 12.4 dB at the identified optimal fibre length of 42 mm, which is actually the switching length.

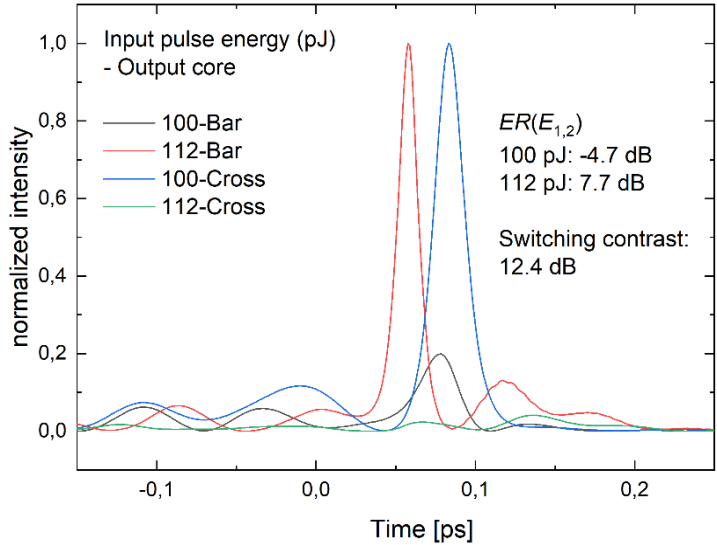

*Figure 7. Normalized-intensity pulse envelopes at the output of each core at the switching length (42 mm) of an all-solid dual-core fibre under excitation by X polarized, 100 fs, 1550 nm pulses. Two envelopes are presented for the both switching energies displaying exchanged dominancy of the bar (red curve) or the cross (blue curve) core with switching contrast of 12.4 dB.*

Figure 7. illustrates the temporal evolution of the intensity of the pulse envelope at 42 mm switching length in the case of 1550 nm excitation wavelength and 100 fs pulse width taken from the four propagation maps in Figure 6. The intensity envelopes are separately normalized for both input pulse energies, thus a value of 1 appears at the peak of the dominant core. The self-trapped soliton in the cross core has longer duration and is delayed in comparison to the bar core self-trapped soliton. Even though, the majority of the energy in the case of the both dominant envelopes is located in time window of 100 fs, bearing potential for data processing rates above 1 Tb/s. Beyond this window, non-negligible energy is observable in the case of the non-dominant envelopes, so the integrated energy in proximity of the centre of the time window should result in even higher switching contrast. Considerable drawback of the obtained results is, that the switching performance is asymmetric with nearly two times higher extinction ratio at the higher pulse energy in comparison to the lower one. It was one of our motivations to perform further numerical analysis, looking for the best switching performance in wider range of excitation wavelengths, pulse widths and propagation distances.

Even though, the presented switching performance is promising, because it predicts better switching contrast than in the case of the air-glass PCF structure [7]. However, now it is ensured in fibre structure with significantly simpler manufacturing technology and with different combination of linear optical fibre characteristics. The nonlinearity coefficient for both fibres is very similar because the same highly nonlinear glass is used for the cores in both cases and the effective mode areas are also at the same level. On the other hand, the simple dual-core all-glass structure has 3 times lower anomalous dispersion and 6 times longer coupling length in the C-band. However, the increase of the wavelength causes increase of the anomalous dispersion and dramatic decrease of coupling length (Figure 3. and 4.). Thus, the solitonic propagation character should depend strongly on the excitation wavelength. According to this expectation we decided to explore spectral range of 1400 - 1800 nm in the context of switching performance to explore the excitation wavelength effects. Similarly, the pulse width effect was studied in the experimentally available range of 75 - 150 fs, since we expected that it may influence also the character of the solitonic propagation (see equations (7) and (8)). We analysed the switching performance for different combinations of excitation wavelength and pulse width and compared to the case of 1550 nm, 100 fs.

In Table 1. we present the switching parameters for different wavelengths at fixed pulse width of 75 fs and at minimal switching energy levels, which ensure self-trapping of the high-order soliton in the cross vs. bar core. In terms of switching energies and length, monotonic dependences were identified. The decrease of the switching length with increasing wavelength is caused mainly by the decrease of the coupling length (see Figure 4.). Considering, that the soliton self-trapping process requires balance between coupling length and soliton compression length, we conclude that the $z_{comp}$, expressed by equation (7), should also decrease with increasing wavelength. The dispersion and the nonlinear parameter which appear in equation (7) possesses weak wavelength dependence with opposite tendencies, therefore their effect is negligible. Thus, the only possibility to decrease $z_{comp}$ for increasing wavelength at fixed pulse width is to increase the peak power - e.g. pulse energy - (see equation (7)). This is in good agreement with the identified wavelength dependence of the switching energies presented in Table 1. According to our studies the optimal wavelength is 1500 nm at 75 fs pulse width, resulting in the switching contrast at level of 17 dB. Consequently, the optimal soliton number supporting the self-trapped solitonic switching is in range of 2-3. The pulses with lower soliton numbers do not induce sufficient peak powers during the compression process to strongly trap the field in one core. On the other hand, the higher *N* causes more complex soliton fission process, which results in leakage of some parts of the field to the opposite core. The results of the wavelength dependence study point out, that the self-trapped solitonic switching is attainable in wider range of input parameters. On the other hand, the optimal switching contrast requires beside the appropriate $z_{comp}$ proper level of the soliton number, as well.

*Table 1. Switching parameters for different excitation wavelengths at fixed pulse width of 75 fs and at minimal switching energy levels, those induce self-trapping of the high-order soliton in the cross (first value) and in the bar core (second value).*

| Excitation wavelength, $\lambda$ [nm] | 1400 | 1500 | 1600 | 1700 | 1800 |
|---|---|---|---|---|---|
| Switching energies, $E_{1,2}$ [pJ] | 17, 22 | 53, 65 | 107, 115 | 215, 232 | 422, 435 |
| Soliton number, *N* | 1.9, 2.2 | 2.4, 2.7 | 2.8, 2.9 | 3.4, 3.6 | 4.3, 4.4 |
| Switching contrast [dB] | 12.5 | **17.2** | 16.1 | 12.4 | 7.2 |
| Switching length [mm] | 62 | 43 | 41 | 32 | 32 |

In the same manner, we analysed the pulse width dependence of the switching performance keeping $\lambda$ fixed. Table 2. presents the values of the switching parameters in the case of $\lambda = 1500$ nm and the pulse width in the range of 75 - 150 fs with a step of 25 fs. The outcome of this analysis is that all parameters express monotonic dependence on the pulse width, with best switching performance at the shortest pulse. The only deviation from this general trend is the slight increase of the switching contrast at pulse width of 125 fs. The switching energies increase almost linearly with increasing pulse width, because the same threshold peak power is required in the all cases for launching the self-trapping process. In this study, the only changing parameter in equation (7) is the pulse width, so its elongation causes increase of $z_{comp}$. According to the example shown in Figure 6., the optimal switching length occurs just above the soliton compression point. Therefore, the increasing $z_{comp}$ is the reason of the monotonic increase of the switching length, although the $L_C$ remains the same due to constant wavelength. The main reason of the switching length change is the increasing soliton number and its optimal level for the high-contrast switching performance was again identified. The same values came out, as in the case of wavelength dependence study, because pulses with widths longer than 75 fs cause rather low switching contrast. We did not consider shorter pulses, because our goal was to ensure switching performance for laser pulses that are available using simple and cheap Er based mode-locked oscillators. Furthermore, the broadband character of shorter pulses would be a disadvantage from the point of view of the dispersion.

*Table 2. Switching parameters for different pulse widths at fixed wavelength of 1500 nm and at minimal switching energy levels, those ensure self-trapping of the high-order soliton in the cross (first value) and in the bar core (second value).*

| Pulse width, $T_{FWHM}$ [fs] | 75 | 100 | 125 | 150 |
|---|---|---|---|---|
| Switching energies, $E_{1,2}$ [pJ] | 53, 65 | 66, 81 | 87, 97 | 116, 126 |
| Soliton number, *N* | 2.4, 2.7 | 3.1, 3.4 | 3.9, 4.2 | 5.0, 5.2 |
| Switching contrast [dB] | **17.2** | 13.1 | 13.6 | 11.8 |
| Switching length [mm] | 43 | 56 | 63 | 80 |

The most important outcome of the two above presented studies is the optimal level of the soliton number to be between 2 and 3. The reason for changing the value of *N* was different in the case of the wavelength and pulse width dependence, even though the same optimal level, which ensures the best switching contrast came out. It is worth mentioning, that the sequence of the identified switching lengths and energies cannot be expressed analytically. The small deviations from the general trends are a natural effect of the soliton self-trapping process, which has a threshold character and experience sudden switching between cores. This bistable process is very sensitive to the all linear and nonlinear propagation mechanisms, therefore they can shift the switching parameters to lower or higher values depending on their actual weight. Therefore, we are satisfied with clear trends of the switching parameters wavelength and pulse width dependences in correspondence with the above presented theoretical considerations.

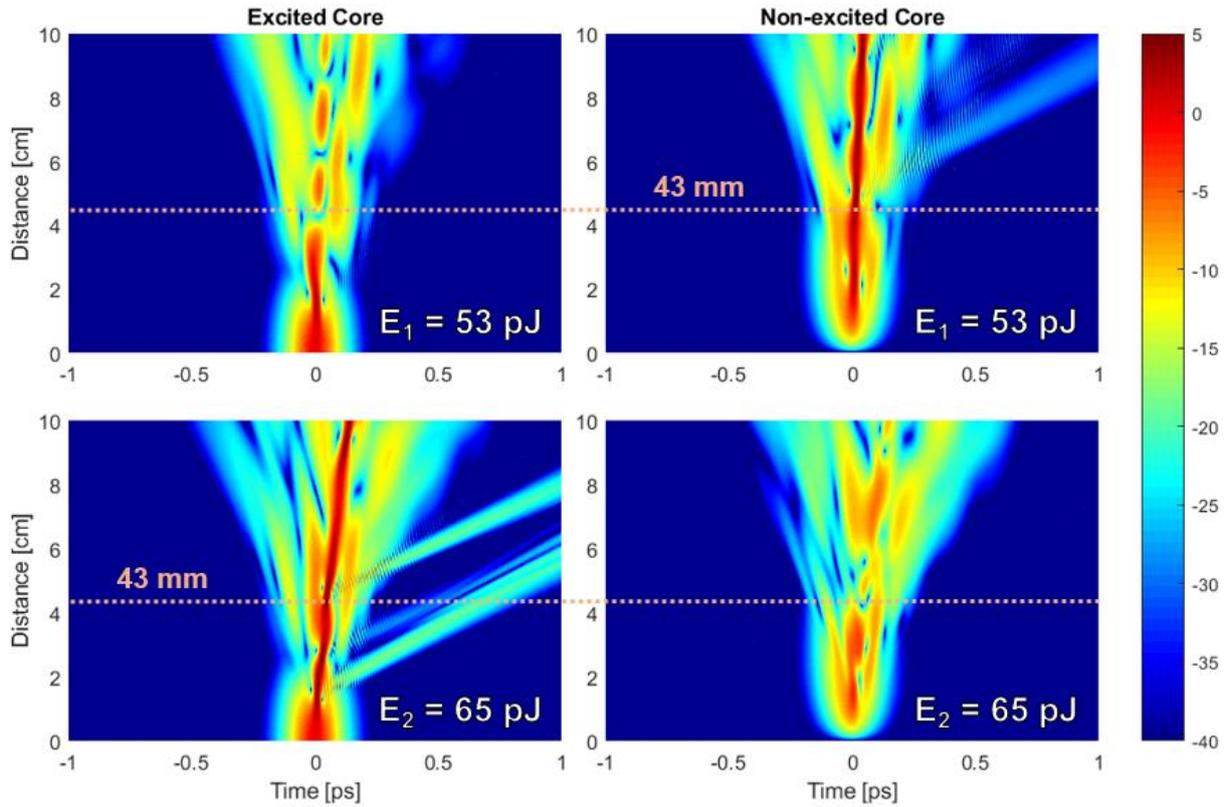

*Figure 8. Numerically generated retarded time frame intensity maps of solitonic dual-core propagation in case of the best switching performance. Excited (bar) core is presented on the left, the non-excited (cross) on the right side, respectively, under excitation by X polarized, 75 fs, 1500 nm pulses for two selected energy levels: 53 pJ (top), 65 pJ (bottom).*

Similar trends as presented in the two tables were predicted in the case of series with other constant pulse widths or excitation wavelengths. Therefore, the combination 1500 nm and 75 fs with the optimal soliton number was not overcome from point of view of switching contrast in the chosen excitation parameter ranges. In Figure 8. the temporal propagation maps corresponding to this best switching performance are presented. When we compare them to the results obtained for combination 1550 nm, 100 fs in Figure 6., the effect of the optimal soliton number is obvious. The less complex propagation character observable in Figure 8. supports higher concentration of energy in the self-trapped dominant lines, which ensures higher switching contrast. Despite of the different excitation parameters the switching length is very similar in the both presented case. This is another confirmation of the trends showed in Table 1,2, the shorter wavelength (1500 nm) causes elongation and the shorter pulse (75 fs) causes shortening of the switching length.

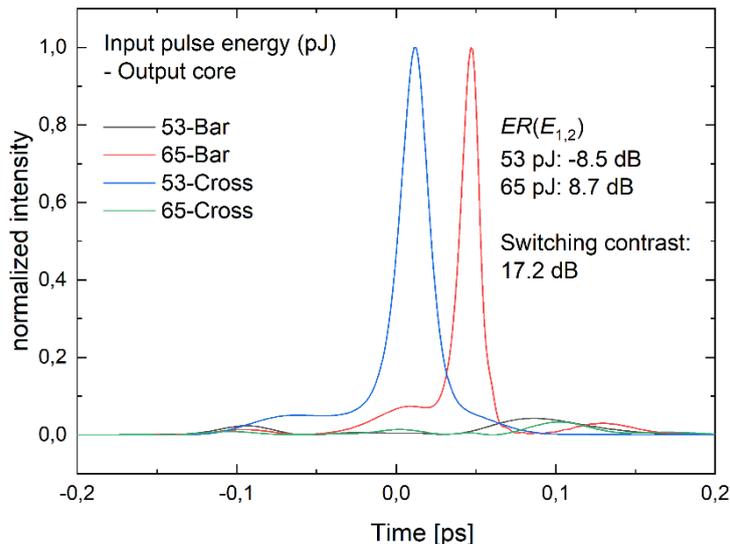

*Figure 9. Normalized-intensity pulse envelopes at the output of each core of the optimized length (43 mm) of an all-solid dual-core fibre in case of X polarized, 75 fs, 1500 nm excitation pulses in case of the best switching performance. Two*

*envelopes are presented for the both switching energies displaying exchanged dominancy of the bar (red curve) or the cross (blue curve) core with the highest switching contrast value of 17.2 dB.*

In Figure 9. the temporal intensity envelopes in the best switching performance case are presented at the optimal 43 mm propagation length. There is similar temporal delay between the dominant pulses in the two different cores, as in the case presented in Figure 7. The reason of the delay is in the propagation history of the self-trapped soliton. The delayed pulse (red curve in Figure 9., blue curve in Figure 7.) has along his path higher average peak powers in comparison to the path of the concurrent self-trapped pulse. The higher average peak powers induced nonlinear refraction index enhancement slows down the pulse in relation to the weaker trapped one in the opposite core. The integral energy switching contrast increased from value 12.4 dB to value 17.2 dB comparing the two presented cases. Importantly, the switching performance in the best switching performance case is symmetric, expresses the same level of dual-core extinction ratio at the both energies (-8.5 vs. 8.7 dB). The switching contrast increases significantly, when we consider the field energy just in the time window of the short self-trapped solitonic pulses. The evaluation of our results is appropriate also from this aspect in the case of interaction of the switched pulses with timely synchronized short optical signals. Similarly, only the time window of the main pulse envelope contributes significantly to a nonlinear signal applying the switched pulses for any type of subsequent nonlinear interaction.

Therefore, we fitted the time envelopes $I(t)$ of the both dominant solitons at the lower and higher energy with sech$^2$ function. The resulted curves $F_{1,2}(t)$ was used to calculate the integral

$$E_j^i = \int F_j(t) I_j^i(t)\, dt, \qquad (10)$$

where the upper index denotes the core and the lower one the energy level, respectively. The obtained four values of the temporally localized energy were used to calculate the *ER* and switching contrast according to equation (9). The same procedure was applied in the case of the first published paper about the air-glass DC PCF [7], expressing integrated energy switching contrast of 11.8 dB and temporally localized one of 34.8 dB. In the case of this new study, the temporally localized switching contrast increased also significantly, expressing value of 46.3 dB. Table 3. reports the general comparison of the switching parameters between the mentioned first study of air-glass DC PCF structure with our recent results. In terms of switching energies, the in-coupled ones are presented, because the in-coupling efficiency could be different in the case of air-glass and all-solid fibre. In the frame of the overview of the all results above, the pulse energy before the in-coupling optics is presented, taking into account the experimentally determined 32 % in-coupling efficiency in the case of the real DC PCF fibre [8]. However, it represents an overestimation of the required pulse energy, because the simple all-solid structure should have significantly higher in-coupling efficiency. Therefore, for comparison purposes between the air-glass PCF and all-solid fibre, it is better to evaluate the in-coupled energies.

In general, the excitation parameters in the both cases are very similar, however, in terms of dispersion and coupling length, the two approaches differs significantly. We conclude that the self-trapped solitonic switching can be realized for various fibres with different linear optical characteristics, so the recent study confirms its application potential. The dispersion optimization lead to a structure with relatively weak coupling in comparison to our previous work, thus it supports switching lengths only at the level of few cm. However, from all other aspects, the new approach has only advantages. The simple dual-core structure requires 6 times lower switching energies and most importantly works at lower soliton number, which enables stronger field trapping after the soliton compression. As we already emphasised, when the value of *N* is above the optimal level of 2 - 3, we obtain lower switching contrast. Therefore, we achieved both integral energy and temporally localised switching contrast enhancement by factor of 30 % in comparison to the air-glass PCF case. The new structure has significant potential for applications due to the simple cladding (avoiding the photonic lattice). Furthermore, the all-solid approach supports better integrability with the standard fibre optics communication systems.

|  | λ [nm] | $T_0$ [fs] | D [ps/nm/km] | $E_{1,2}$ [pJ] | N | $z_{comp}$ [mm] | $L_c$ [mm] | sw. length [mm] | sw. contrast [dB] | T sw. contr. [dB] |
|---|---|---|---|---|---|---|---|---|---|---|
| Air-glass PCF | 1550 | 100 | 169 | 115, 124 | 4.1, 4.3 | 3.4, 3.2 | 3.7 | 10.3 | 11.8 | 34.8 |
| All-solid simple | 1500 | 75 | 56.1 | 17, 21 | 2.4, 2.7 | 13.7, 11.7 | 33 | 43 | 17.2 | 46.3 |

*Table 3. Comparison of the switching parameters in the case of earlier studied air-glass DC PCF structure [7] and in the case of simple all-solid core-cladding structure. The new structure supports the switching performance at the optimal soliton number and it enables higher switching contrasts at simultaneous lower switching energy requirement.*

From the point of view of potential experimental realization, it is necessary to address two further questions: how robust is the presented best switching performance against the fluctuations of the input pulse energy (peak power) and of the structural dimensions of the fibre? In order to answer the first question, the nonlinear simulations were performed for the values of switching energies close to the optimal ones (53 pJ, 65 pJ) considering the same optimized dual-core all-solid structure with excitation parameters of 1500 nm and 75 fs. The results are presented in Table 4. The extensive study revealed that the upper energy level is very robust against fluctuations of the pulse energy even at the level of ± 5 %. It means that energy pairs such 53 pJ, 68 pJ or 53 pJ, 62 pJ result in the same high contrast switching performance at the level of 17 dB. The soliton remains to be strongly trapped in the bar core in the whole energy range of 62 – 68 pJ. The lower energy level is also robust. However, it is centered around 50 pJ, which means that in the energy range of 47 – 53 pJ the soliton is strongly trapped in the cross core. The combination of 50 pJ with the higher energy level limiting values of 62 pJ and 68 pJ gives again the switching contrasts at the level of 17 dB, however the optimal switching length is longer and is around 58 mm. Only in the case of the lowest value of 47 pJ we obtained decrease of the switching contrasts to the value of 13 dB at switching lengths in the range of 55 – 60 mm.

| Switching energies, $E_{1,2}$ [pJ] | 47, 62 | 47, 65 | 47, 68 | 50, 62 | 50, 65 | 50, 68 | 53, 62 | 53, 65 | 53, 68 |
|---|---|---|---|---|---|---|---|---|---|
| Switching contrast [dB] | 12.4 | 12.8 | 13.2 | 16.3 | 16.7 | 16.8 | 16.6 | 17.2 | 17 |
| Switching length [mm] | 60.0 | 56.7 | 55.7 | 58.2 | 58.3 | 57.7 | 44.8 | 43.0 | 43.8 |

*Table 4. Effect of input energy fluctuation at level of ± 5 % on switching contrast and switching length in the conditions of the best-found switching performance.*

However, the standard femtosecond Er doped mode-locked fibre oscillators exhibit amplitude rms noise at the level of 0.5 %, so such fluctuations will not compromise the switching parameters predicted in the case of the best switching performance. The aspect of the structural dimension fluctuations along the fibre axis is even less crucial due to the short optimal fibre length of about 5 cm. The monitoring of the diameter of soft glass fibres during the drawing process along comparable lengths displays fluctuations at the level of 1 %, which causes negligible change of linear and nonlinear properties of the fibre.

## 5. SUMMARY

In conclusion, the capability of an all-solid high-contrast dual-core fibre for all-optical switching applications was introduced in this paper. A novel switching concept, developed by us earlier, based on soliton self-trapping is numerically studied in the case of such a special dual-core fibre. We optimized the fibre material and the microstructure geometry to make fabrication process simpler and ensure comparable values of coupling and soliton compression length when sub-100 pJ ultrafast pulses are applied. The developed and already available soft glass lead silicate – borosilicate pair is well matched thermally and expresses refraction index contrast at the level of 0.4 in the near infrared. Nonlinear propagation of $sech^2$ pulses was studied numerically in the optimized fibre structure at different excitation wavelengths (1400 - 1800 nm) and at different pulse widths (75 - 150 fs). The comparison of the switching parameters in the case of various excitation wavelength, pulse width combinations revealed the effect of the underlying propagation mechanisms on them. Most importantly, the optimal level of soliton number in range of 2 - 3 were identified. The best parameter combination of 1500 nm, 75 fs, causing the proper soliton number in conditions of self-trapped solitonic switching at 43 mm fibre length, requires only in-coupled pulse energies at the level of 20 pJ. Furthermore, the robustness of the suggested approach against the fluctuations of the input pulse energies was confirmed. The accompanied switching contrasts in the case of integral and temporally localized field energies were at level of 17 dB and 46 dB, respectively. These outcomes represent significant progress in comparison to the previously studied air-glass DC PCF structure at simultaneous 6 times lower in-coupled pulse energies. Additionally, the simpler manufacturing process enables to work more intensively on the development of an appropriate fibre. In summary, the presented study introduces a realistic way towards the first demonstration of highly effective, applicable, soliton based all-optical switching using a simple optical fibre.

**ACKNOWLEDGEMENTS**

This work was supported by the Foundation for Polish Science co-financed by the European Union under the European Regional Development Fund [TEAM TECH/2016-1/1], by National Science Centre, Poland [project No. 2016/23/P/ST7/02233 under POLONEZ program which has received funding from the European Union's Horizon 2020 research and innovation program under the Marie Skłodowska-Curie grant agreement No 665778], by Slovak R&D Agency under the contracts No. SK-AT-2017-0026, APVV-14-0716 and by Austrian Agency OeAD under contract number SK 02_2018.